\documentstyle[prl,aps,multicol,graphicx]{revtex}

\newcommand{\be}{\begin{equation}} \newcommand{\ee}{\end{equation}}

\newcommand{\ba}{\begin{eqnarray}}
\newcommand{\ea}{\end{eqnarray}}

\newcommand{\bn}{{\bf n}}

\begin{document}
\draft

\title{Cluster Percolation and Chiral Phase Transition}

\author{Matteo Beccaria$^{a, b}$, Antonio Moro$^{a, b}$}

\address{ ${}^a$ Dipartimento di Fisica dell'Universit\`a di Lecce,
I-73100, Italy,\\ ${}^b$ Istituto Nazionale di Fisica Nucleare - INFN,
Sezione di Lecce }

\maketitle

\begin{abstract}
The Meron Cluster algorithm solves the sign problem in a class of interacting
fermion lattice models with a chiral phase transition. Within this framework, 
we study the geometrical
features of the clusters built by the algorithm, that suggest the occurrence 
of a generalized percolating phase transition at the chiral
critical temperature in close analogy with Fortuin-Kasteleyn percolation in spin models.
\end{abstract}


\begin{multicols}{2}
\narrowtext

A fundamental difficulty in the Monte Carlo study of fermion lattice models
is the sign problem due to the fluctuating sign of the statistical
weight of fermion configurations~\cite{Linden}. 
Recently, the Meron Cluster algorithm (MCA)~\cite{Meron,Chandr} has
been proposed  
as an effective solution to the sign problem in a class of interacting models. 
In particular, here, we focus on a $2+1$ dimensional
model with a second order phase transition 
associated to the dynamical breaking of a discrete chiral symmetry~\cite{2+1}.

Like all cluster algorithms for lattice models, MCA
defines clusters of sites used as effective non-local 
degrees of freedom 
to update configurations without critical slowing down.
For a given observable, the sign problem is cured by 
restricting the Monte Carlo sampling to specific topological sectors 
that give contributions not canceling in pairs due to the fermion sign.
The relevant topological charge is the so-called meron number that we shall
define later.
The rules that assure convergence to the correct 
Boltzmann equilibrium distribution determine a well defined cluster dynamics.
From the study of lattice spin models we know that this artificial
dynamics can be surprisingly rich; in fact, experience in that field
suggests the existence of a purely geometrical phase transition 
concerning the algorithm clusters and underlying the physical 
thermal transition~\cite{FK,Sokal,Fortunato}.

The simplest example of such a scenario is the 2D Ising model
where clusters can be defined in a natural way as sets of nearest neighboring 
aligned spins and admit a physical interpretation 
as real ordered domains.
At the thermal transition temperature these clusters 
percolate~\cite{ConiglioNappi}, but since the  critical exponents 
do not coincide with the thermal ones~\cite{Sykes}, a
complete equivalence 
of the two transitions cannot be claimed.
In three dimensions, the comparison is even worse and  also 
the two critical temperatures are slightly different~\cite{Krumb}. 
To find a geometrical transition occurring at the thermal 
critical point with the same critical exponents, generalized clusters
must be defined~\cite{ConiglioKlein}, like the Fortuin-Kasteleyn bond clusters 
and their extensions~\cite{FK,Sokal}. 
The equivalence between the thermal phase transition and a suitable
percolative process can then be extended to models with continuous
rotational or gauge invariance~\cite{Fortunato}.

Similar investigations lack for fermionic models with 
sign problems because cluster algorithms have not been available until MCA. 
Here, for the first time, we aim to the identification of a geometrical 
transition in the MCA
dynamics and look for critical phenomena defined in terms of cluster shapes.
On the other hand, we must keep into account at least the global 
configuration signs
because they are the only memory of the fact that the model is fermionic and allow to 
tell it from its bosonized counterpart free of sign problems.
Besides, apart from the sign problem, 
the rules to build clusters in MCA are not precisely the same as for 
Fortuin-Kasteleyn clusters and the existence of a transition is non trivial.

The model we study describes relativistic staggered fermions, hopping on a
$2+1$ dimensional lattice with $L^2$ spatial sites, described
in~\cite{2+1}.
The Hamiltonian is
\ba
 H &=& \sum_x\sum_{i=1,2} \left\{
\eta_{x,i}(c^\dagger_x c_{x+\hat\i} + h.c.) + \right. \\ &+& \left.
G\left(n_x-\frac 1 2 \right)\left(n_{x+\hat\i}-\frac 1 2 \right)
\right\} ,
\ea
where $n_x=c_x^\dagger c_x$ is the occupation number at site $x$,
$\eta_{x,i}$ are the Kawamoto-Smit phases $\eta_{x,1}=1$,
$\eta_{x,2}=(-1)^{x_1}$ and the operators $c$, $c^\dagger$ obey
standard anti-commutation relations $\{c_x, c_y\}=0$, $\{c_x,
c_y^\dagger \}=\delta_{x,y}$. In the following we shall consider the
case $G=1$ and adopt periodic boundary conditions.

The partition function $\mbox{Tr}\ e^{-\beta H}$ can be computed by
Trotter splitting that maps the quantum model to a 
statistical system on a $2+1$ dimensional lattice with $L^2\times T$ 
sites. The temporal lattice spacing is $\varepsilon = 4\beta/T$. 
The limit $\varepsilon\to 0$ must be taken at fixed $\beta$ 
with $T\to \infty$.  In practice, we shall present results 
obtained at the fixed value $T=40$ where $\beta$ can cover the
transition point with reasonably small $\varepsilon$~\cite{2+1}.  

Each configuration is specified by the occupation numbers 
$\bn=\{n_{x,t}\}$ and carries a sign $\sigma(\bn)=\pm 1$, source of
the sign-problem. To update a configuration, 
sites are clustered according to definite rules depending on $\beta$ and discussed 
in details in~\cite{Chandr}.
Each cluster is then independently flipped: with probability $1/2$ 
we apply the global transformation $n_{x,t}\to 1-n_{x,t}$ to 
all of its sites. Clusters whose flip 
changes $\sigma(\bn)$ are defined merons.

The chiral phase transition can be analyzed by studying the asymptotic
volume dependence of the susceptibility $\chi$. 
Defining the chiral condensate in the 
configuration $\bn =\{n_{x,t}\}$ as
\be
Z(\bn) = \frac\varepsilon 4\sum_{x,t}(-1)^{x_1+x_2}\left(n_{x,t}-\frac
1 2 \right),
\ee 
the chiral susceptibility $\chi$ is given by
\be
 \chi = \frac{1}{\beta
L^2}\frac{\langle (Z(\bn)^2)\ \sigma(\bn)\rangle}
{\langle\sigma(\bn)\rangle} .
\ee
An improved estimator of $\chi$ free of sign problems can be built by 
writing $Z(\bn)$ as a sum over clusters $Z(\bn)=\sum_C Z_C$ and taking
its average over $2^{N_C}$ possible flips, where $N_C$ is the
number of the clusters. Then,
$\chi$ gets contributions from sectors with meron number $N=0,2$:
\be
 \chi = \frac{1}{\beta L^2}
\frac{\langle \sum_C Z_C^2\ \delta_{N,0}+2 |Z_{C_1}Z_{C_2}|\ 
\delta_{N,2}\rangle}{\langle\delta_{N,0}\rangle} , 
\ee
where, for $N=2$, $C_1$, $C_2$ are the two merons.  

From numerical simulations, apart from rather small scaling violations, $\chi$ obeys the 
Finite Size Scaling (FSS) law 
$ \chi = L^\gamma f_\chi(L(\beta-\beta_{th})) $
with the exponent $\gamma=7/4$, characteristic of the 2D Ising
universality class~\cite{Kogut}.
Numerical simulations locate the chiral $\beta$ at
$\beta_{th} = 2.43(1)$~\cite{2+1}(the
subscript ``{\em th}'' stands for ``{\em thermal}'').

To detect a possible purely geometrical transition, we study
quantities $Q$ that depend only on the cluster shape and not on their
internal occupation numbers. 
Since cluster flips do not change $Q$ and meron flips change the sign of 
$\sigma(\bn)$, the improved estimator of $Q$ is
\be
 \frac{\langle Q(\bn)\
\sigma(\bn)\rangle}{\langle\sigma(\bn)\rangle} = \frac{\langle Q(\bn)\
\delta_{N,0}\rangle}{\langle\delta_{N,0}\rangle} \equiv \mbox{\bf
E}_0(Q), 
\ee
that is the average restricted to the zero meron sector.

\begin{figure}[htb]
\centerline{\includegraphics*[width=6cm,angle=-90]{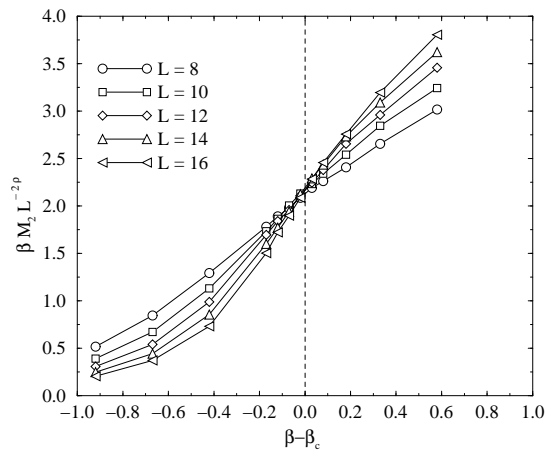}}
\caption{Data crossing for $M_2$ with different $L$. By virtue of the 
scaling law Eq.~(\ref{trivial}), at $\beta=\beta_c$ the quantity $\beta M_2 L^{-2\rho}$ is
independent on $L$.}
\label{fig:cross}
\end{figure}

\begin{figure}[htb]
\centerline{\includegraphics*[width=6cm,angle=-90]{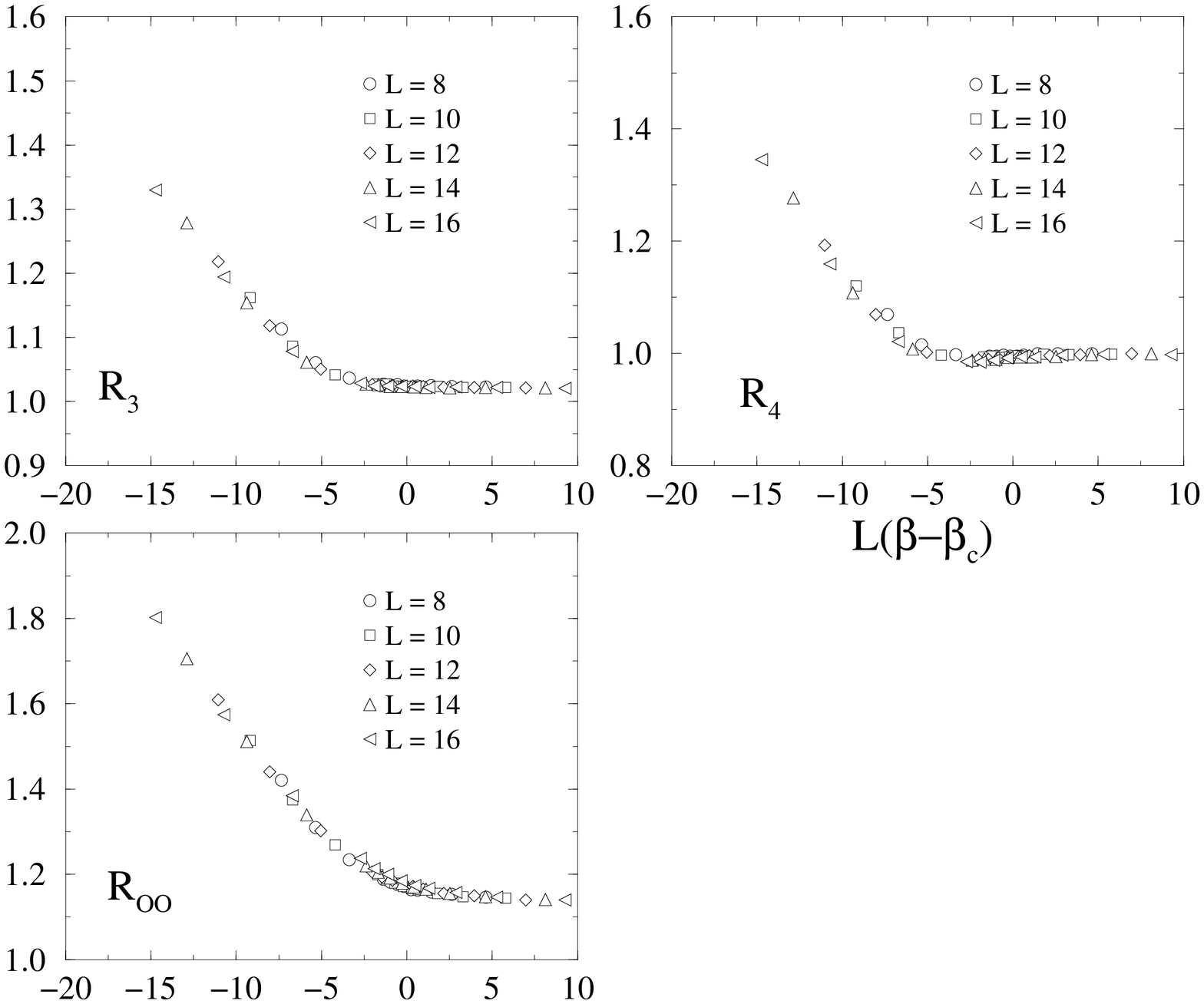}}
\caption{Ratios $R_n$ and $R_\infty$. $M_n^{1/n}$
is well approximated by the contribution of the largest cluster beyond
the critical point.}
\label{fig:ratios}
\end{figure}

The simplest set of geometrical 
quantities that we can study are the moments 
$M_n = \mbox{\bf E}_0\left(\sum_C |C|^n\right)$ of 
the normalized cluster size $|C| = \mbox{Vol}(C)/T$ where $\mbox{Vol}(C)$
is the number of sites in $C$.
\begin{figure}[htb]
\centerline{\includegraphics*[width=6cm,angle=-90]{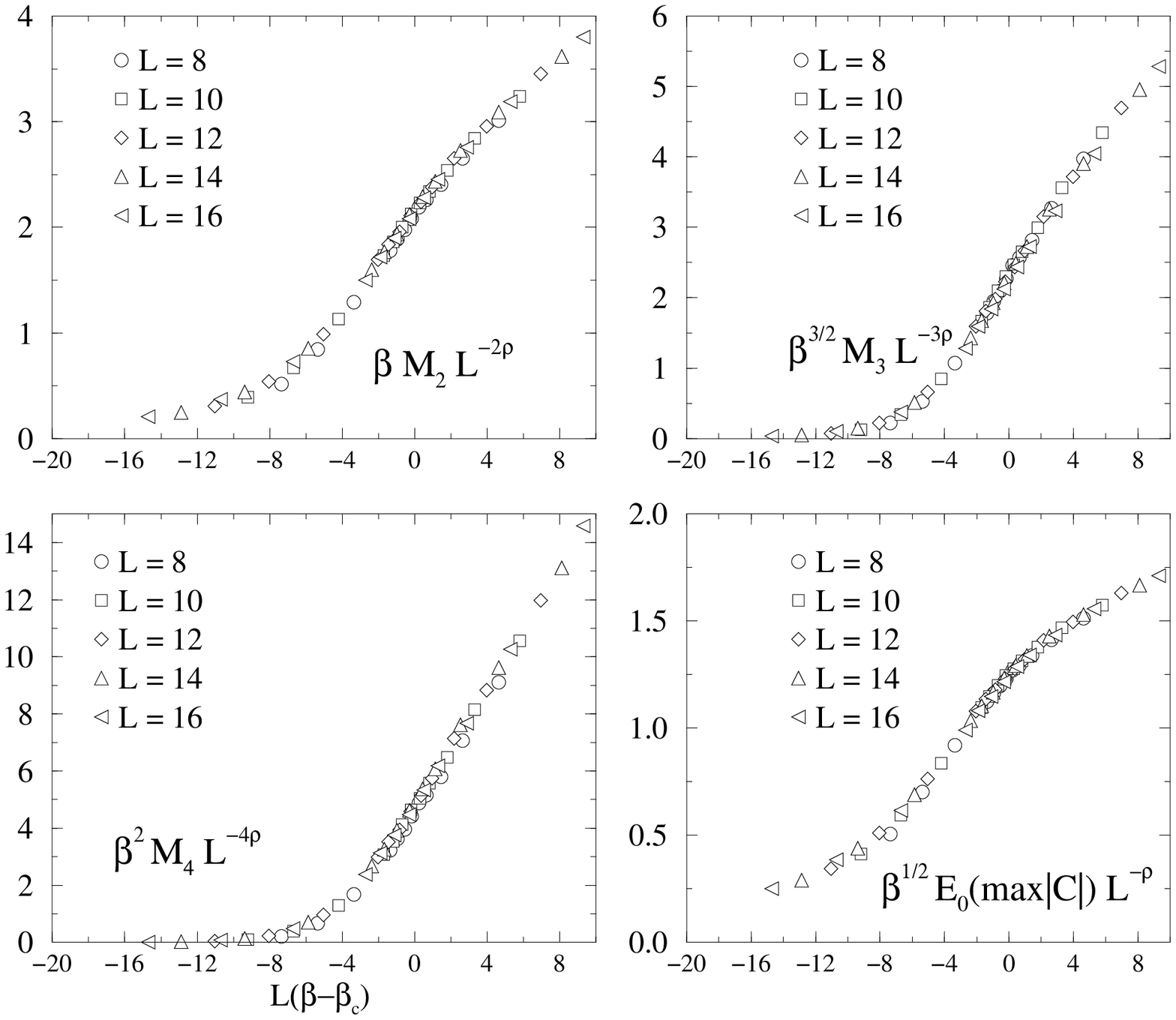}}
\caption{FSS analysis 
of $M_n$ and $\mbox{\bf E}_0(\max |C|)$.} 
\label{fig:scaling}
\end{figure}
\noindent 
We perform simulations to compute numerically $\{M_n\}_{n=2,3,4}$ 
and also $\mbox{\bf E}_0(\max |C|)$. 
We work on lattices with 
$L=8$, $10$, $12$, $14$, $16$ in the range $1.5 < \beta < 3.0$
and presented about $1.5\cdot 10^5$ measures per point. 
Fig.~(\ref{fig:scaling}) shows the numerical data supporting as a first result
the remarkable validity of the empirical scaling relations:
\ba
\label{trivial}
\beta^{n/2}\ M_n &=& 
L^{n\rho}\ f_n(y), \quad n\ge 2, \\
\beta^{1/2}\ \mbox{\bf E}_0(\max |C|) &=& 
L^{\rho}\ h(y) \nonumber
\ea
in terms of the scaling variable $y=L(\beta-\beta_c)$.
This scaling behavior defines an order parameter of the
geometrical phase transition.
The results for the exponent $\rho$ and the critical $\beta_c$ are
$\rho = 1.71(5)$ and $\beta_c = 2.42(5)$.
Within errors, $\rho$ is consistent 
with the exact 2D Ising exact value $\gamma = 7/4$. 
We also find $\beta_c \simeq \beta_{th}$ with an accuracy that  can 
be appreciated by looking at Fig.~(\ref{fig:cross}).
\noindent
The ratios $R_n = M_2^{1/2}/M_n^{1/n}$ for $n=3, 4$
and $R_\infty = M_2^{1/2}/\mbox{\bf E}_0(\max |C|)$ are shown in 
Fig.~(\ref{fig:ratios}).
\noindent
The plots strongly indicate that $R_n$ are independent on 
$L$ at fixed $y$ in agreement with Eq.~(\ref{trivial}).

\begin{figure}[htb]
\centerline{\includegraphics*[width=6cm,angle=-90]{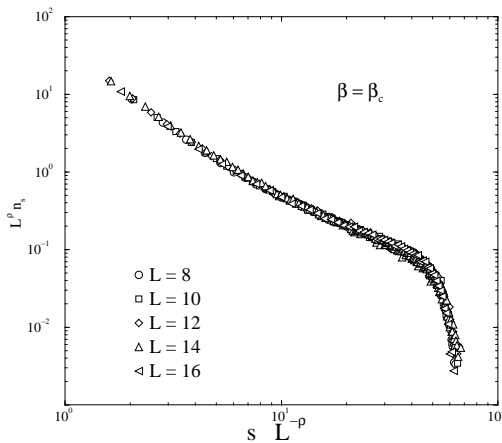}}
\caption{FSS plot of $n_s$. 
The curves are obtained after averaging over blocks of 10 subsequent
sizes.
In the region with $L^{\rho} < s < 40\ L^\rho$, $n_s$ decreases roughly
algebraically; beyond the {\em knee} at $s \simeq 40\ L^{\rho}$  the
distribution 
falls down quickly. The critical region is dominated by
the clusters at the right edge of the plot.}
\label{fig:ns}
\end{figure}

It can be checked that, at $\beta=\beta_c$, $M_n$ receives the main
contribution from a small
set of large clusters growing like $L^\rho$, and
the fact that the ratios $R_n(y) \rightarrow 1$  
as $y$ grows, simply means that this contribution is more and more
dominant.
We briefly summarize this behavior by saying that the clusters are
{\em percolating}.
Following~\cite{Stauffer}, further information on the critical 
ensemble can be obtained by studying the cluster distribution
\be
n_s = \mbox{\bf E}_0(\#\ \mbox{number of clusters $C$ with
$\mbox{Vol}(C) = s$}) . 
\ee
In Fig.~(\ref{fig:ns}) we show that
the simple law 
\be
\label{ns}
n_s = L^{-\rho} f\left(\frac{s}{L^\rho}\right),
\ee
is well satisfied for $s/L^{\rho} \gg 1$ .
\noindent

The critical cluster distribution $n_s$ decreases not faster 
than algebraically with $s$ until $s\simeq 40\ L^\rho$ where
the final large cluster tail is reached and $n_s$ falls down quickly.

If one takes into account that the leading contribution to
$\{M_n\}_{n\ge 2}$ actually comes from the region where $s/L^{\rho} \gg 1$,
observing that $M_n = \sum_s s^n n_s$ and using Eq.~(\ref{ns}) we obtain
a $L$-dependence consistent with Eq.~(\ref{trivial}).  

There are many small clusters and one can check that the 
normalized number of clusters $\beta\ \mbox{\bf E}_0(N_C)/L^2$ depends 
mildly on $L$ for a wide range of $\beta$ and $L$. 

\begin{figure}[htb]
\centerline{\includegraphics*[width=6cm]{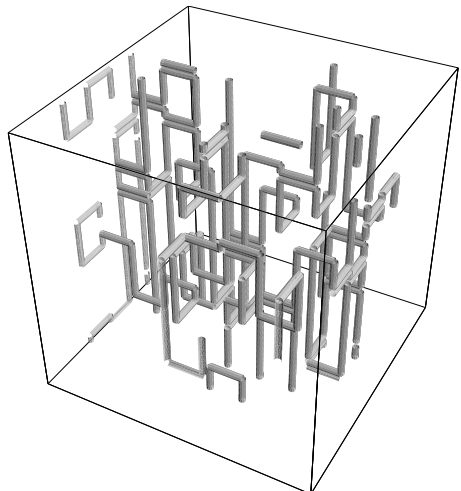}}
\caption{A typical largest link configuration
for the $8^2\times 40$ lattice at
$\beta=\beta_c$.}
\label{fig:tube2.50}
\end{figure}

It is interesting to analyze what happens to the typical 
cluster configurations when $\beta$ is gradually increased
toward the critical point. 
At small $\beta$, almost all bonds that build the clusters 
are set in the temporal direction. In physical terms the fermions hop
 from the initial 
positions to neighboring ones with small probability.
Sites simply tend to cluster in $L^2$ straight vertical lines
with $T$ sites. As $\beta\to\beta_c$, the vertical clusters 
start merging and breaking and form complicated structures 
with large dispersion in the cluster size distribution.  
This process can be seen in Fig.~(\ref{fig:breaking}): 
\begin{figure}[htb]
\centerline{\includegraphics*[width=6cm]{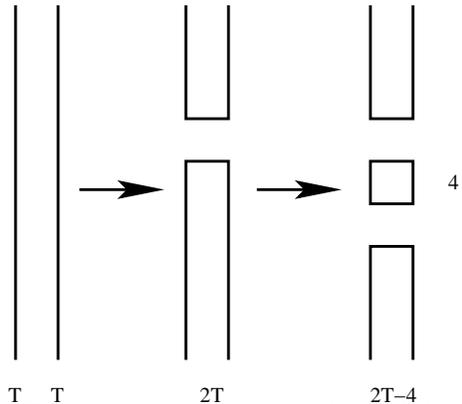}}
\caption{Two elementary clusters with
equal dimension $T$, first merge into a single cluster, then break
apart in two clusters with very different dimension.} 
\label{fig:breaking}
\end{figure}
\noindent
two vertical clusters with $\mbox{Vol}(C)=T$ undergo a two step process 
allowed by the cluster rules~\cite{Meron}. In the end, they give rise to a 
large cluster with $\mbox{Vol}(C)=2T-4$ and a small one with 4 sites.
After many processes of this kind we find a few large clusters 
surrounded by a gas of smaller clusters. In Fig.~(
\ref{fig:tube2.50}) 
\noindent
we show a typical largest cluster obtained at $\beta = 2.5$ on the 
$8^2\times 40 $ lattice.

Geometrical quantities that can measure this dispersion effect are the cumulants
$\{G_n\}_{n\ge 1}$ 
of the cluster size distribution defined as
\be
 \sum_{n\ge
1}\frac{\lambda^n}{n!} G_n = \mbox{\bf E}_0\left\{ \log\left(
\sum_{n\ge 0} \frac{\lambda^n}{n!}\frac{1}{N_C}\sum_C |C|^n
\right) \right\} .
\ee

In our study we do not consider the cumulant $G_1$,
whose behavior is determined by the 
the contributions from small clusters, with $s\ll L^\rho$.
  
The next cumulant is the variance
\be
G_2 = \mbox{\bf
E}_0\left(\frac{1}{N_C}\sum_C |C|^2 -\frac{1}{N_C^2}(\sum_C
|C|)^2\right) .
\ee
Our data support a FSS law of the form 
\be
\label{G2}
\beta G_2 = L^{\rho'} g_2(y),
\ee
with $\rho' = 1.51(4)$, which is compatible with $2\gamma-2$.
This exponent can be explained taking account that, analogously to the
moment $M_2$, the contributions
from small clusters are negligible and that, within errors,
$\sum_C |C|^2 \sim L^{2\gamma}$ and $N_C \sim L^2$ .

\begin{figure}[htb]
\centerline{\includegraphics*[width=6cm,angle=-90]{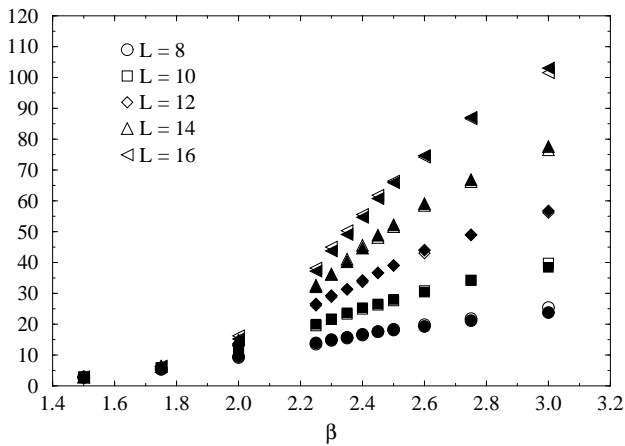}}
\caption{Comparison between $\chi$ (empty symbols) and 
$0.15\cdot (\beta G_2)^{7/6}$ (full symbols).}
\label{fig:ChiG2Ratio}
\end{figure}

As we remarked above, when $\beta \rightarrow 0$, the clusters are
simple vertical lines and $G_2 \rightarrow 0$ like $\chi$ does.
This fact suggests that it could be interesting to study a possible relationship
between these quantities at different $\beta$ values. 
We remark that, in principle, $G_2$ and $\chi$ have different topological
origins: $\chi$ is calculated on zero and two merons sector, 
while $G_2$ only on zero-meron sector. 
Nevertheless, we find that the following empirical relation 
\be
\label{ChiG2}
\chi \simeq 0.15(1) \cdot (\beta\ G_2)^{\frac 1 2\frac{\gamma}{\gamma-1}},
\ee
holds for a wide range of parameters $(L, \beta)$ as shown in 
Fig.~(\ref{fig:ChiG2Ratio}).
\noindent
In practice, in the region that we have explored, 
the ratio $\chi/(\beta G_2)^{7/6}$ 
is consistent, within errors, with a constant.
This result signals 
an unexpected correlation between different topological sectors.
In particular, Eq.~(\ref{ChiG2}) could be relevant in the construction
of a purely geometrical definition of $\chi$.
A FSS study of the higher cumulants $G_3$ and $G_4$ 
shows similar scaling laws, but with exponents that are not in simple 
relation with $\gamma$ and should in principle be matched to the 
anomalous dimensions of higher operators 
in the 2D Ising universality class.

To conclude, we have examined the critical behavior of the clusters
that arise in the application of the Meron algorithm to a fermion
model in $2+1$ dimensions. We have found simple FSS laws
that we have explained in terms of a percolative process occurring at the 
chiral critical temperature. Our data support 
the results $\rho \simeq \gamma$, $\rho' \simeq 2\gamma-2$,  as well
as the empirical relation Eq.~(\ref{ChiG2}) that shows a 
close correlation between physical and geometrical quantities.

We acknowledge S. Chandrasekharan, K. Holland and
U.J. Wiese for useful discussions about the Meron-Cluster Algorithm and
its applications. Financial support from INFN, IS-RM42 is also
acknowledged.

\end{multicols}
\end{document}